\documentclass[12pt]{iopart}
\usepackage{graphicx}

\begin{document}

\title[Analysis of experimental quantum gates]{Analysis of an experimental quantum logic gate \\
by complementary classical operations}

\author{Holger F. Hofmann}

\address{Graduate School of Advanced Sciences of Matter, Hiroshima University\\ Kagamiyama 1-3-1, Higashi Hiroshima 739-8530, Japan
\\hofmann@hiroshima-u.ac.jp
}

\author{Ryo Okamoto}

\address{Research Institute for Electronic Science,
Hokkaido University
\\
Sapporo 060-0812, Japan
}

\author{Shigeki Takeuchi}

\address{Research Institute for Electronic Science,
Hokkaido University
\\
Sapporo 060-0812, Japan
}

\begin{abstract}
Quantum logic gates can perform calculations much more efficiently
than their classical counterparts. However, the level of control
needed to obtain a reliable quantum operation is correspondingly
higher. In order to evaluate the performance of experimental
quantum gates, it is therefore necessary to identify the
essential features that indicate quantum coherent operation.
In this paper, we show that an efficient characterization of
an experimental device can be obtained by investigating the
classical logic operations on a pair of complementary basis sets.
It is then possible to obtain reliable predictions about the
quantum coherent operations of the gate such as entanglement
generation and Bell state discrimination even without
performing these operations directly.
\end{abstract}


\section{Introduction: Quantum Computation Processes}

Within recent years, quantum computation has become a well
established field of research in both experimental and theoretical
physics. At the heart of this field is the notion that the
highly entangled correlations of many-particle quantum systems
could be used as a tool to efficiently solve problems of equally
challenging complexity. In order to convert a quantum system
from a mere object of observation into a problem solving tool,
it is necessary to establish a nearly complete control over
quantum processes at the microscopic level.

In  close analogy to conventional computation, the method of
establishing this high level of control over large quantum
systems is to assemble the quantum systems from the smallest
possible element offered by quantum theory, the two level
system. For obvious reasons, this two level system is then
referred to as a quantum bit or qubit. However, its physical
properties are better visualized by the analogy with the three dimensional spin of a spin-1/2 system.
In fact, one possible explanation for the efficiency of quantum
computation is the fact that the possibilities of rotating a
spin are infinite, while a classical bit can only be flipped.
Intriguingly, quantum mechanics smoothly connects these seemingly
contradictory aspects of reality in a single consistent theory.

In principle, it is possible to construct a universal quantum
computer using only local spin rotations and a single well-defined
interaction \cite{NieC4}. One such well-defined interaction between
two qubits is the quantum controlled-NOT gate. When observed in the
computational basis (usually associated with the $z$-component in
the spin analogy), this gate performs a classical controlled-NOT
logic operation. However, it is completely quantum coherent, so
its actual performance is far more complex than that of its
classical namesake.

Since the successful realization of a quantum controlled-NOT
would enable universal quantum computation, a significant
amount of experimental effort has been devoted to this goal.
(see ref.~\cite{Tur95} to \cite{Oka05} for examples.)
However, experimental realizations are never identical
to the ideal device described by theory. In order to demonstrate
that an experimental device really performs the intended function,
it needs to be tested. For classical logic gates, such a test is
straightforward, since the number of possible operations
is finite. But for operations on qubits, the possibility of
arbitrarily small phase shifts implies that the number of
possible quantum coherent operations is in principle infinite.
Therefore, the experimental test of a quantum gate requires a
somewhat deeper understanding of the essential features of
general quantum operations. In particular, we need to go beyond the
rather fuzzy image of quantum coherence and the associated
``parallelism'' of quantum superpositions, towards a more specific
approach based on the observable features of quantum devices.

In this review, we briefly introduce the proper theoretical
description of experimental quantum processes. We then show that
the essential features of a quantum process can be defined in
terms of only two complementary operations \cite{Hof05} and
derive estimates for the quantum process fidelity and the
entanglement capabilities based on the corresponding complementary
classical fidelities.
Finally, we present a recently realized optical controlled-NOT
gate \cite{Oka05} and show how information about the actual
device performance can be obtained from the experimental data.

\section{Theoretical Description of Noisy Quantum Operations}

Ideally, a quantum operation can be represented by a unitary
operator $\hat{U}_0$ acting on the input state
$\mid \psi_{\mbox{in}} \rangle$ in the $d$-dimensional Hilbert space
of the quantum system. Since all quantum states can be expanded in
terms of a complete orthogonal basis set $\{ \mid n \rangle \}$, the
effect of the unitary operation on an arbitrary input state is completely
defined by its effects on such a set of $d$ basis states,
\begin{equation}
\label{eq:U}
\hat{U}_0 \mid n \rangle = \mid f_n \rangle.
\end{equation}
Because the operation is unitary, the output states
$\{ \mid f_n \rangle \}$ also form an orthogonal basis set.
The quantum operation is thus completely deterministic and leaves
no room for unpredictable errors. In particular, it should be
noted that the phases of the states $\mid f_n \rangle$ are also
defined by eq. (\ref{eq:U}), so that the unitary transformation
actually defines much more than the transformation of an eigenvalue
$n$ to a corresponding eigenvalue $f_n$.

Obviously, it is very difficult to realize a nearly deterministic
error free quantum operation experimentally. The idealized description
given by a single unitary operation $\hat{U}_0$ is therefore not normally
sufficient to describe noisy experimental processes.
Instead, we have to assume that the actual process $\hat{A}_m$ acting
on the input state $\mid \psi_{\mbox{in}} \rangle$ may fluctuate
randomly and is not necessarily unitary.
If the probability distribution over possible processes $\hat{A}_m$
is given by $p_m$, the output state is described by
a mixed state density matrix \cite{NieC8},
\begin{equation}
\label{eq:noise}
\hat{\rho}_{\mbox{out}} = \sum_m p_m
\hat{A}_m \mid \psi_{\mbox{in}} \rangle
\langle \psi_{\mbox{in}} \mid \hat{A}_m^\dagger.
\end{equation}
In general, any reproducible quantum process can be represented
in such a form. However, if the precise source of errors is
unknown, it is not possible to identify a unique set of operations
$\hat{A}_m$. For the experimental evaluation of quantum processes,
it is therefore more useful to find a representation that
does not depend on the specific error syndromes $\hat{A}_m$.

It is in fact possible to express any noisy process in a $d$-dimensional
Hilbert space in terms of an orthogonal set of $d^2$ operators
${\Lambda_i}$ by considering the $d \times d$ matrices representing
the operators as vectors in a $d^2$-dimensional vector space
\cite{NieC8,Mahler}. An ideal process can then be expressed as
\begin{equation}
\hat{A}_m = \sum_i c_i \hat{\Lambda}_i,
\end{equation}
and any noisy process $E$ can be described by a process matrix
with elements $\chi_{ij}$, so that
\begin{equation}
\hat{\rho}_{\mbox{out}} = E(\hat{\rho}_{\mbox{in}})
= \sum_{i,j} \chi_{ij} \hat{\Lambda}_i
\hat{\rho}_{\mbox{in}} \hat{\Lambda}_j^\dagger.
\end{equation}
Each process can thus be decomposed into a finite set of orthogonal
processes $\{ \hat{\Lambda}_i \}$, and the complete process is then
defined by its process matrix elements $\chi_{ij}$.

In principle, the complete $d^4=16^{N}$ process matrix elements can
always be evaluated by measuring the output statistics of a sufficient number of non-orthogonal input states \cite{NieC8}. This approach, called quantum process tomography, treats the quantum process as a black box,
requiring no further assumptions about the intended
process itself. In order to test a specific quantum operation however,
it may be more useful to formulate the process matrix in terms of
basis processes $\{ \hat{\Lambda}_i \}$ that are close to experimentally observable error syndromes of the device. As we show in the following,
it is then possible to obtain useful information about the device performance without an abstract analysis of the huge amount of data
required for complete quantum process tomography.

\section{Classification of Quantum Errors}

For qubits, each error can be expanded in terms of products of
errors acting on a single qubit, and the single qubit errors can be
expressed in terms of the identity $I$ and the three Pauli matrices,
$X$, $Y$, $Z$ \cite{NieC8,Mahler}. Using the spin analogy, these errors
correspond to spin flips (rotations of 180 degrees) around the
$x$-, $y$- and $z$-axis, respectively. An $N$-qubit system is thus
characterized by the $N$-qubit identity $\hat{F}_0$ and $4^N-1$ spin
flip errors, $\hat{F}_i$.

If the intended operation is $\hat{U}_0$, errors will be detected
by comparing the output qubit statistics with the ideal operation.
It is therefore useful to characterize the errors with reference
to $\hat{U}_0$ as output errors $\hat{U}_i=\hat{F}_i \hat{U}_0$.
The noisy process is then described by
\begin{equation}
\label{eq:pm}
E(\hat{\rho}_{\mbox{in}})
= \sum_{i,j} \chi_{ij} \hat{U}_i
\hat{\rho}_{\mbox{in}} \hat{U}_j^\dagger,
\end{equation}
where the diagonal elements $\chi_{ii}$ of the process matrix now
correspond to the distribution of spin-flip errors in the output.
It is thus possible to identify experimentally observed output
errors directly with a group of theoretical error syndromes
and their corresponding process matrix elements.

We now have a convenient mathematical form for the representation
of errors in a quantum operation. However, we still need to determine
the errors experimentally, so it is necessary to consider
the observable effects of the errors for a given set of output states.
Since most quantum information processes are formulated in the
computational basis defined by the eigenvalues of $Z$, it is useful
to start by considering an operation which produces the $Z$ basis
states in the output, $\mid f_n \rangle =
\mid Z_1,Z_2, \ldots \rangle$. In the $Z$ basis, the operators $X$
and $Y$ represents bit flips, and the operators $I$ and $Z$ preserve
the qubit value. $Y$ and $Z$ also change the phase relation between
the qubit states, but this phase change cannot be observed in the
$Z$ basis. Therefore, it is not possible to distinguish $I$ from $Z$
or $Y$ from $X$ when the output is measured in the $Z$ basis.

Most importantly, a quantum device that always produces the correct
$Z$ output may still have phase errors that destroy the quantum
coherence between the outcomes. In fact, there are a total of $2^N$
mutually orthogonal operations consistent with the correct $Z$ basis
output of an $N$ qubit operation, defined by assigning either the
identity $I$ or the phase flip $Z$ to each qubit.
In order to detect these errors, it is necessary to perform an
operation that is sensitive to $Z$-errors in the output.
Since the $Z$-errors represent spin rotations around the $Z$-axis,
this is most naturally achieved by using a complementary
set of inputs $\mid k^\prime \rangle$ that result in $X$ basis
outputs, $\mid g^\prime_k \rangle = \mid X_1,X_2, \ldots \rangle$.
In this basis, the $Z$-errors show up as bit flips, so that
all error syndromes will show up either in the $Z$-operation or
in the $X$-operation.

Whether an experimental quantum process really performs the
intended quantum coherent operation $\hat{U}_0$ can therefore be
tested efficiently by observing the classical logic operations
in the computational $Z$ basis and the complementary classical
logic operation in the $X$ basis. If both operations are performed
with high fidelity, the device will also perform any other quantum
coherent operation reliably well.

\section{Evaluation of Device Performance}


We have now seen that only an ideal error free quantum process
can produce correct outputs in both the $Z$- and the $X$ basis.
However, experimental processes will usually show errors in
both operations. To evaluate these errors, it is necessary to
introduce measures that do not depend on the choice of $Z$ and
$X$ outputs, but are equally valid for any kind of quantum
coherent operation.

One such measure immediately suggests itself from the formulation
of the process matrix in eq.(\ref{eq:pm}). Since the matrix element
$\chi_{00}$ represents the probability of the correct quantum
operation $\hat{U}_0$ (as opposed to the probabilities of the
errors $\hat{U}_i$ given by $\chi_{ii}$), it seems natural to
identify $\chi_{00}$ with the quantum process fidelity $F_{qp}$.
In fact, this definition is now widely used to evaluate
quantum processes based on the full process matrix obtained by
quantum tomography \cite{NieC8}. However, it is not immediately
clear from eq.(\ref{eq:pm}) how the matrix element $\chi_{00}=F_{qp}$
relates to the individual fidelities observed for specific input
states $\mid \psi_{\mbox{in}} \rangle$.
To get a more intuitive understanding of quantum process fidelity,
it is therefore useful to know that $F_{qp}$ can also be defined
operationally, as the fidelity that would be obtained by applying
the process to one part of a maximally entangled pair of $N$-level
systems. If the maximally entangled state is given by $\mid
\mbox{E}_{\mbox{\tiny max}} \rangle_{AB}$ and the processes $E_A$ and acts only on
system $A$, $F_{qp}$ can then be defined as
\begin{equation}
\label{eq:Fqp}
F_{qp} = \langle \mbox{E}_{\mbox{\tiny max}} \mid (\hat{U}_0^\dagger \otimes I)_{AB}
E_A(\mid \mbox{E}_{\mbox{\tiny max}} \rangle \langle
\mbox{E}_{\mbox{\tiny max}} \mid)
(\hat{U}_0 \otimes I)_{AB} \mid
\mbox{E}_{\mbox{\tiny max}} \rangle = \chi_{0,0}.
\end{equation}
The application of a quantum process to one part of an entangled
pair is thus sensitive to all possible errors $\hat{U}_i$.

An even better intuitive understanding of the process fidelity can be
obtained by considering the relation between eq.(\ref{eq:Fqp}) and the
fidelity expected for a randomly selected local input state
$\mid \psi_{\mbox{in}} \rangle_A$ in system $A$.
In fact, any such state can be
prepared from$\mid \mbox{E}_{\mbox{\tiny max}} \rangle_{AB}$ by simply performing
a local measurement on system $B$. It is then possible to derive
a relation between the process fidelity $F_{qp}$, and the average
fidelity $\bar{F}$, defined as the probability of obtaining the
correct output averaged over all possible input states \cite{Hor99},
\begin{equation}
\bar{F} = \frac{F_{qp}\, d + 1}{d + 1}.
\end{equation}
In the light of the present error analysis, this relation can now
be understood in terms of the sensitivity of the (local) input states
$\mid \psi_{\mbox{in}} \rangle_A$ to the different errors $\hat{U}_i$.
Specifically, the discussion of $X$ and $Z$ output errors above
has shown that these operations are insensitive to exactly $d-1$
out of the $d^2-1$ possible errors $\hat{U}_i$. We can conjecture
that any input state $\mid \psi_{\mbox{in}} \rangle_A$ is
insensitive to
a fraction of $1/(1+d)$ of all possible errors. Therefore, a process
fidelity of $F_{qp} = 0$ results in an average fidelity $\bar{F}$
of exactly $1/(1+d)$, representing the probability of finding input
states that are insensitive to the errors of the operation.

After having convinced ourselves of the usefulness of the quantum
process fidelity $F_{qp}$ as a measure of the general device
performance, we can now return to the task of determining this
measure from a limited number of test measurements.
For this purpose, we need to define the classical fidelities
of the two complementary operations resulting in $Z$ or in
$X$ output states. These fidelities are directly obtained by
averaging over the $d$ probabilities of measuring the correct output
state $\mid~f_n\rangle = \mid~Z_1,Z_2,\ldots\rangle$
or $\mid~g^\prime_k\rangle = \mid~X_1,X_2,\ldots\rangle$
after applying the quantum process $E$ to an input of
$\mid n \rangle = \hat{U}_0^\dagger \mid f_n \rangle$
or $\mid k^\prime \rangle = \hat{U}_0^\dagger
\mid g^\prime_k \rangle$,
\begin{eqnarray}
\label{eq:Fclass}
F_Z &=& \frac{1}{d} \sum_n \langle f_n \mid
E(\mid n \rangle \langle n \mid) \mid f_n \rangle
\nonumber \\
F_X &=& \frac{1}{d} \sum_k \langle g^\prime_k \mid
E(\mid k^\prime \rangle \langle k^\prime \mid) \mid g^\prime_k \rangle.
\end{eqnarray}
By applying the definition of these classical fidelities to the
process matrix representation in eq.(\ref{eq:pm}), we find that the
classical fidelities $F_Z$ and $F_X$ are given by sums of diagonal
elements $\chi_{ii}$. For simplicity, we will now label the errors
$i$ according to their effects on $Z$ and $X$ outputs. Each error
is then identified by a pair of bit flip patterns, $i=j_z j_x$,
where $j_{z}=0$ ($j_{x}=0$) indicates no error in the $Z$ ($X$)
basis outputs. The diagonal elements contributing
to $F_Z$ and $F_X$ are then given by
\begin{eqnarray}
\label{eq:Fchi}
F_Z &=& \chi_{00,00} + \sum_{j \neq 0} \chi_{0j,0j}
\nonumber \\
F_X &=& \chi_{00,00} + \sum_{j \neq 0} \chi_{j0,j0}.
\end{eqnarray}
Each classical fidelity thus includes the process fidelity
$\chi_{00,00}=F_{qp}$ and a different set of error probabilities,
$\chi_{0j,0j}$ for $F_Z$ and $\chi_{j0,j0}$ for $F_X$.

Since the diagonal elements of the process matrix must add
up to one, it is possible to define an additional relation
between the total number of errors and the process fidelity,
\begin{equation}
\label{eq:sum}
\chi_{00,00} = 1 - \sum_{l \neq 0} \chi_{0l,0l}
- \sum_{m \neq 0} \chi_{m0,m0}
- \sum_{l,m \neq 0} \chi_{ml,ml}.
\end{equation}
With this relation, it is possible to express the process fidelity
$\chi_{00,00}=F_{qp}$ in terms of a sum of the classical complementary
fidelities and the probabilities $\chi_{ml,ml}$ for errors observed in both
operations ($m\neq 0$ and $l\neq 0$),
\begin{equation}
\label{eq:Frel}
F_{qp} = F_Z + F_X -1 + \sum_{l,m \neq 0} \chi_{ml,ml}.
\end{equation}
This is a significant result, since it provides a quantitative
lower limit of the process fidelity $F_{qp}$ using only the classical
fidelities obtained from two times $d$ orthogonal input states.
In addition, eq. (\ref{eq:Fchi}) provides an upper limit by showing
that the process fidelity is always lower than the classical fidelities.
The process fidelity $F_{qp}$ is therefore limited to an interval of
\cite{Hof05}
\begin{equation}
\label{eq:estimate}
F_Z + F_X -1 \leq F_{qp} \leq \mbox{Min}\{F_Z, F_X\}
\end{equation}
defined by the experimental results for the fidelities $F_Z$ and $F_X$
of the two complementary classical operations observed in the
$X$ and the $Z$ basis.

\section{Measures for the Non-Locality of a Quantum Process}

Up to now, we did not discuss any specific properties of the operation
$\hat{U}_0$, and all of the arguments above also apply to the problem
of transmitting a string of qubits unchanged (intended operation
$\hat{U}_0=I$). In fact, the original definitions of fidelities,
used e.g. in ref. \cite{Hor99}, actually derive from this problem
of characterizing quantum channels.
However, the purpose of quantum computation is the manipulation of
entanglement. It is therefore essential that the operations are capable
of generating and discriminating various kinds of entangled states.

In particular, the generation of entanglement is commonly recognized as
a key feature of genuine quantum operations, and has therefore been
used extensively as an experimental criterion for the successful
implementation of quantum gates. The most widely
used figure of merit is the entanglement capability $C$, defined as
the maximal amount of entanglement that the gate can generate from
local inputs \cite{Poy97}.
However, it is usually not easy to determine the
amount of entanglement of an experimentally generated state.
Instead, the most simple experimental approach is to estimate
the minimal entanglement necessary to obtain an experimentally
observed correlation average expressed by so-called entanglement
witnesses \cite{Hor96,Ter00,Lew00,Tot05}. In the present context,
the most useful
entanglement witnesses are the ones constructed from the projection
on the intended entangled state $\mid E_{\mbox{out}} \rangle$,
\begin{equation}
\hat{W} = \frac{1}{1-b} \left(
\mid E_{\mbox{out}} \rangle \langle E_{\mbox{out}} \mid - b
\right),
\end{equation}
where $b$ is the maximal fidelity of $\mid E_{\mbox{out}} \rangle$
for non-entangled states ($b=1/M$ for $M \times M$ entanglement).
It is then possible to derive a measure of the entanglement
capability directly from the fidelities of entanglement generation.
Specifically, if the ideal quantum process $\hat{U}_0$ is capable of generating a maximally entangled state from local inputs, the
minimal entanglement capability of an experimental realization with
process fidelity $F_{qp}$ is simply given by
\begin{equation}
\label{eq:Cest}
C \geq \frac{1}{1-b} \left(F_{qp} - b\right),
\end{equation}
since $F_{qp}$ is the minimal probability of obtaining the correct
output state.

In addition to the generation of entanglement, quantum gates can
also perform the reverse operation of converting entangled
inputs into local outputs. At first sight, it may not be clear
why this is useful, since decoherence and local measurements
appear to have the same effect. However, only non-local quantum
operations can decode the quantum information encoded in a
entangled states by transforming orthogonal entangled inputs into
orthogonal local states. The measure of non-locality for the
``disentanglement''  of entangled states is therefore the
capability of distinguishing orthogonal entangled states.
Since orthogonal entangled states are often referred
to as Bell states, a device with this capability is also known as
a Bell analyzer. It may therefore be useful to define another measure
of non-locality to characterize the operation of such Bell analyzers.

In the following, we define the entanglement discrimination $D$
using the fidelities $F_i$ of the operations
\begin{equation}
\hat{U}_0 \mid E_i \rangle = \mid L_i \rangle,
\end{equation}
where $\{\mid E_{1/2} \rangle\}$ are two orthogonal entangled inputs
and $\{\mid L_{1/2} \rangle\}$ are the corresponding orthogonal
local outputs. An operation that cannot distinguish the two input
states generates the same random output for both inputs, so the
maximal average fidelity $F_{\mbox{av.}}=(F_1+F_2)/2$ is $1/2$.
We therefore define the entanglement discrimination as
$D= 2 F_{\mbox{av.}}-1$. Since $F_{\mbox{av.}}$ must be greater than
or equal to $F_{qp}$, we obtain an estimate of the entanglement
discrimination of
\begin{equation}
\label{eq:Dest}
D \geq 2 F_{qp} - 1
\end{equation}
from the process fidelity $F_{qp}$.

It may be worth noting that this estimate corresponds to the
entanglement capability estimate for $M=2$ ($b=1/2$).
In the following analysis of a quantum controlled-NOT operation,
it can be seen that this similarity arises from the time-reversal
symmetry of entanglement generation and entanglement discrimination.
It is therefore possible to generalize the definition of
entanglement discrimination to the case of distinguishing $M$
orthogonal $M \times M$ entangled states, so that the two
estimates become equal. In the present context, however, the
simple definition of pair discrimination will be sufficient.

\section{Analysis of an Optical Quantum Controlled-NOT}

\begin{figure}[ht]
\begin{center}
\scalebox{0.45}[0.45]{\includegraphics{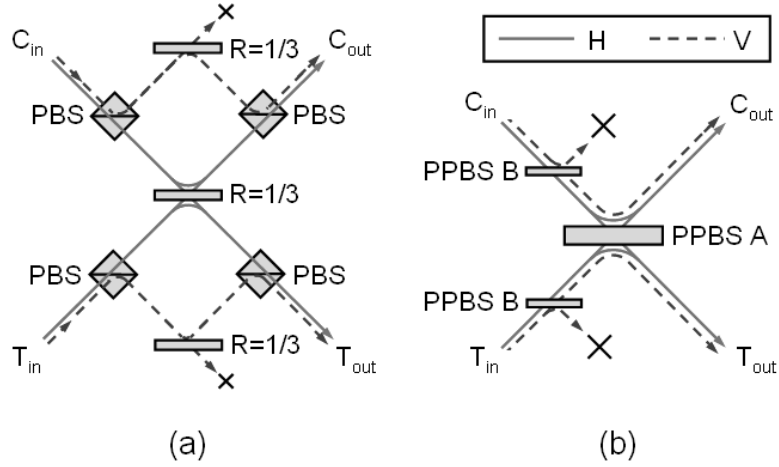}}
\end{center}
\caption{\label{setup} Schematics of the optical quantum
controlled-NOT gate. (a) shows the original proposal
using the two photon interaction at a beam splitter of
reflectivity 1/3 and (b) shows the recently developed
compact realization using partially polarizing beam
splitters (PPBS).}
\end{figure}

We have now analyzed the theoretical possibilities of errors
in quantum devices and their general effects on entanglement
generation and discrimination. Based on this foundation, we can
now proceed to evaluate experimental data from an actual
quantum process realized in the laboratory.
The device we will consider is a quantum controlled-NOT based
on linear optics and post-selection \cite{Hof01,Ral01}. In this
device, a non-linear interaction between two photonic qubits
is achieved by interference between the two
photon reflection and the two photon transmission components at
a central beam splitter of reflectivity $1/3$, as shown in
fig. \ref{setup} (a). Recently, several groups succeeded in
developing a very compact version of this device, where each
photonic qubit follows only a single optical path and the
interaction is realized at a partially polarizing beam splitter
(PPBS) of reflectivity 1/3 for horizontally ($H$) polarized
light and reflectivity 1 for vertically ($V$) polarized light
\cite{Lan05,Kie05,Oka05}. The schematic setup of this device is
shown in fig. \ref{setup} (b). Details of the specific experimental
setup developed by us can be found in ref. \cite{Oka05}.

As in most experiments using photons as qubits, the input
photon pairs were generated by spontaneous type II parametric
downconversion using a beta barium borate (BBO) crystal.
The crystal was pumped by an argon ion laser at a wavelength
of 351.1 nm, generating photon pairs in orthogonal polarization
at a wavelength of 702.2 nm. The photon polarization (that is,
the local states of the qubits) of these photon pairs
was then controlled by half wave plates to achieve the desired
input states. After the controlled-NOT operation at the PPBS,
the output polarizations of the photons was detected by a
standard setup using another set of half wave plates, polarization
beam splitters, and single photon counters (SPCM-AQ-FC, Perkin Elmer).

In the context of our optical quantum controlled-NOT gate,
the computational $Z$ basis and the complementary $X$ basis
are defined in terms of the linear polarizations of
the photons. Using the horizontal and vertical polarization
states, $\mid H \rangle$ and $\mid V \rangle$, the corresponding
basis states of the control qubit $C$ and the target qubit $T$ read
\begin{equation}
\begin{array}{rcccrcc}
\mid 0_Z \rangle_C &=& \mid V \rangle &\hspace{1cm}&
\mid 0_Z \rangle_T &=& \frac{1}{\sqrt{2}}(\mid H \rangle + \mid V \rangle)
\\
\mid 1_Z \rangle_C &=& \mid H \rangle &&
\mid 1_Z \rangle_T &=& \frac{1}{\sqrt{2}}(\mid H \rangle - \mid V \rangle)
\\
\mid 0_X \rangle_C &=& \frac{1}{\sqrt{2}}(\mid H \rangle + \mid V \rangle)
&& \mid 0_X \rangle_T &=& \mid V \rangle
\\
\mid 1_X \rangle_C &=& \frac{1}{\sqrt{2}}(\mid H \rangle - \mid V \rangle)
&& \mid 1_X \rangle_T &=& \mid H \rangle.
\end{array}
\end{equation}
The ideal operation $\hat{U}_0$ performed by the quantum controlled-NOT
in the $Z$ basis is a classical controlled-NOT gate. In the complementary
$X$ basis, the roles of target and control are exchanged, and the
classical operation observed is a reversed controlled-NOT gate
\cite{Hof05b}. To demonstrate the successful implementation of a quantum
controlled-NOT, it is therefore sufficient to show that the device
can perform both classical controlled-NOT operations.


\begin{table}[ht]
\caption{\label{table1}Measurement results for the controlled-NOT operation in the $Z$ basis (a) and the complementary reverse controlled-NOT operation in the $X$ basis (b). Bra notation defines outputs and ket notation defines inputs. Numbers in bold face are used for the fidelities of the
correct gate operations.}
\begin{center}
{\large
\begin{tabular}{l|cccc}
(a) & \hspace{0.15cm} $\langle 0_z0_z|$ \hspace{0.15cm}
    & \hspace{0.15cm} $\langle 0_z1_z|$ \hspace{0.15cm}
    & \hspace{0.15cm} $\langle 1_z0_z|$ \hspace{0.15cm}
    & \hspace{0.15cm} $\langle 1_z1_z|$ \hspace{0.15cm}
\\ \hline
$|0_z0_z \rangle$ & {\bf 0.898} & 0.031 & 0.061 & 0.011
\\
$|0_z1_z \rangle$ & 0.021 & {\bf 0.885} & 0.006 & 0.088
\\
$|1_z0_z \rangle$ & 0.064 & 0.027 & 0.099 & {\bf 0.810}
\\
$|1_z1_z \rangle$ & 0.031 & 0.096 & {\bf 0.819} & 0.054
\\
\multicolumn{5}{c}{}
\\
(b) & \hspace{0.15cm} $\langle 0_x0_x|$ \hspace{0.15cm}
    & \hspace{0.15cm} $\langle 0_x1_x|$ \hspace{0.15cm}
    & \hspace{0.15cm} $\langle 1_x0_x|$ \hspace{0.15cm}
    & \hspace{0.15cm} $\langle 1_x1_x|$ \hspace{0.15cm}
\\ \hline
$|0_x0_x \rangle$ & {\bf 0.854} & 0.044 & 0.063 & 0.039
\\
$|0_x1_x \rangle$ & 0.013 & 0.099 & 0.013 & {\bf 0.874}
\\
$|1_x0_x \rangle$ & 0.050 & 0.021 & {\bf 0.871} & 0.058
\\
$|1_x1_x \rangle$ & 0.019 & {\bf 0.870} & 0.040 & 0.071
\end{tabular}
}
\end{center}
\end{table}
Table \ref{table1} shows the experimental results obtained from
our device, as first reported in ref. \cite{Oka05}. The individual
output fidelities are given by the numbers in bold face.
The averages of these values define the classical fidelities
$F_z$ and $F_x$ according to eq.(\ref{eq:Fclass}).
We obtain $F_z=0.85$ and $F_x=0.87$.
Without analyzing any further details, we can now apply
eq.(\ref{eq:estimate}) to determine that the process fidelity
of our gate must be in the range of
\begin{equation}
0.72 \leq F_{qp} \leq 0.85.
\end{equation}
Using eqs.(\ref{eq:Cest}) and (\ref{eq:Dest}), we can then show
that our gate has a minimal entanglement capability $C$ and a
minimal entanglement discrimination of
\begin{equation}
C \geq 0.44 \hspace{0.5cm} \mbox{and} \hspace{0.5cm}
D \geq 0.44.
\end{equation}
We can therefore conclude that our gate can generate and discriminate entanglemed states, based only on the classical fidelities of local input-output relations.

\section{Error Models for the Experimental Device}


A more detailed analysis of our gate is possible if we include
the error probabilities of the two complementary operations
shown in table \ref{table1}. The output errors
$\eta_{x/z} (j_{x/z})$ of each
classical operation can be classified according to the bit flip
errors in the output, using 0 for no error, C for a control flip,
T for a target flip, and B for a flip of both output bits.
The averaged errors from table \ref{table1} then read
\begin{eqnarray}
\eta_z(\mbox{C}) = 0.052 &\hspace{1cm}& \eta_x(\mbox{C})=0.071
\nonumber \\
\eta_z(\mbox{T}) = 0.051 &\hspace{1cm}& \eta_x(\mbox{T})=0.034
\nonumber \\
\eta_z(\mbox{B}) = 0.044 &\hspace{1cm}& \eta_x(\mbox{B})=0.028.
\end{eqnarray}
Likewise, the error operators $\hat{F}_{i}$ can be defined by the
corresponding errors in $Z$ and in $X$, using
$i=\{$ 00, C0, T0, B0, 0C, CC, TC, BC, 0T, CT, TT,
BT, 0B, CB, TB, BB $\}$ to define the output errors
$\hat{F}_i=\{$ II, XI, IX, XX, ZI, YI, ZX, YX, IZ, XZ, IY,
XY, ZZ, YZ, ZY, YY $\}$. Each of the six classical
errors $\eta_{x/z} (j_{x/z})$ can then be identified with a
sum over four diagonal elements $\chi_{ii}$ of the process matrix,
as shown in table \ref{table2}.

\begin{table}[ht]
\caption{\label{table2} Sum relation between experimentally observed
errors and process matrix elements (*=0,C,T,B).}
\begin{center}
{\large
\begin{tabular}{c|cccc|c}
 $\chi_{i,i}$ & *0 & *C & *T & *B & sum \\
\hline
 0* & $\chi_{\mbox{\small 00,00}}$ & $\chi_{\mbox{\small 0C,0C}}$
    & $\chi_{\mbox{\small 0T,0T}}$ & $\chi_{\mbox{\small 0B,0B}}$ & $0.853$ \\
 C* & $\chi_{\mbox{\small C0,C0}}$ & $\chi_{\mbox{\small CC,CC}}$
    & $\chi_{\mbox{\small CT,CT}}$ & $\chi_{\mbox{\small CB,CB}}$ & $0.052$ \\
 T* & $\chi_{\mbox{\small T0,T0}}$ & $\chi_{\mbox{\small TC,TC}}$
    & $\chi_{\mbox{\small TT,TT}}$ & $\chi_{\mbox{\small TB,TB}}$ & $0.051$ \\
 B* & $\chi_{\mbox{\small B0,B0}}$ & $\chi_{\mbox{\small BC,BC}}$
    & $\chi_{\mbox{\small BT,BT}}$ & $\chi_{\mbox{\small BB,BB}}$ & $0.044$ \\
\hline
sum & $0.867$        & $0.071$        & $0.034$        & $0.028$        & $1.000$
\end{tabular}
}
\end{center}
\end{table}

Even though it is not possible to identify the precise values of the
diagonal elements, the sum rules and the positivity of the matrix
elements $\chi_{ii}$ impose strong limitations on the possible
error distributions. For example, the minimal process fidelity
$F_{qp}$ is only obtained if all $\chi_{ii}$ representing errors
in both $X$ and $Z$ are zero. The remaining diagonal elements of the
process matrix are then given directly by the experimentally
observed errors, as shown in table \ref{table3}.
\begin{table}[ht]
\caption{\label{table3} Diagonal elements of the process matrix
for the minimal process fidelity of $F_{qp}=0.72$.}
\begin{center}
{\large
\begin{tabular}{c|cccc|c}
 $\chi_{i,i}$ &
  \hspace{0.3cm} *0  \hspace{0.2cm} &
  \hspace{0.25cm} *C  \hspace{0.25cm} &
  \hspace{0.25cm} *T  \hspace{0.25cm} &
  \hspace{0.2cm} *B  \hspace{0.3cm} & sum \\
\hline
 0* & 0.720 & 0.071
    & 0.034 & 0.028 & 0.853 \\
 C* & 0.052 & 0.000
    & 0.000 & 0.000 & 0.052 \\
 T* & 0.051 & 0.000
    & 0.000 & 0.000 & 0.051 \\
 B* & 0.044 & 0.000
    & 0.000 & 0.000 & 0.044 \\
\hline
sum & 0.867        & 0.071        & 0.034        & 0.028        & 1.000
\end{tabular}
}
\end{center}
\end{table}

The estimates of the diagonal elements of the process matrix can
now be used to derive estimates for the fidelities of operations
other than the observed controlled-NOTs in the $Z$ and $X$ basis.
In particular, the available data allows more detailed predictions
about processes where one qubit
is in the $Z$ basis and the other is in the $X$ basis.
As will be shown in the following, the correct
minimal classical fidelities for these operation on the $ZX$
or $XZ$ eigenstates can in fact be determined by using the process
matrix elements $\chi_{00,00}$ obtained from the minimal process
fidelity estimate given in table \ref{table3}.

The most simple example is the operation on $ZX$ eigenstates
where the control qubit input is in a $Z$ state and the target
qubit input is in an $X$ state.
Since these states are eigenstates of the ideal quantum
controlled-NOT operator $U_0$, the ideal gate performs the identity
operation on these $ZX$ inputs. We can now estimate the minimal fidelity
of this identity operation from table \ref{table1} by identifying
the output errors that preserve $ZX$,
$\hat{F}_i=\{II,ZI,IX,ZX\}$. The classical fidelity $F_I$ of
the identity operation is therefore given by
\begin{equation}
F_I=\chi_{\mbox{00,00}} +\chi_{\mbox{0C,0C}}
   +\chi_{\mbox{T0,T0}} +\chi_{\mbox{TC,TC}}.
\end{equation}
This fidelity can be minimized by associating the errors
$\eta_z(\mbox{C})$ with $\chi_{\mbox{\small C0,C0}}$,
$\eta_z(\mbox{B})$ with $\chi_{\mbox{\small B0,B0}}$,
$\eta_x(\mbox{T})$ with $\chi_{\mbox{\small 0T,0T}}$, and
$\eta_x(\mbox{B})$ with $\chi_{\mbox{\small 0B,0B}}$.
The errors changing the control qubit in $Z$ and the target
bit in $X$ then contribute separately to the total errors
in the identity operation on $ZX$ states, and the minimal
fidelity is given by
\begin{equation}
F_I \geq
 1-(\eta_z(\mbox{C})+\eta_z(\mbox{B})+\eta_x(\mbox{T})+\eta_x(\mbox{B}))
= 0.842.
\end{equation}
As mentioned above, this result is consistent with the distribution
of process matrix elements shown in table \ref{table3}, indicating
that the assumption of a minimal process fidelity of $F_{qp}=F_Z+F_X-1$
also implies a minimal fidelity $F_I$ for the identity operation.

Next, we can analyze the entanglement generation from inputs
in $XZ$ eigenstates. The ideal operation $\hat{U}_0$ generates
maximally entangled two qubit Bell states from each of the
possible $XZ$ inputs. We can therefore derive an estimate of
the entanglement capability $C$ from the fidelity $F_C$ of this
operation. Again, we first identify the errors
output errors that preserve the output states. In this case,
these errors are $\hat{F}_i=\{II,XX,YY,ZZ\}$,
corresponding to an entanglement generation fidelity of
\begin{equation}
F_C=\chi_{\mbox{00,00}} +\chi_{\mbox{B0,B0}}
   +\chi_{\mbox{BB,BB}} +\chi_{\mbox{0B,0B}}.
\end{equation}
Like the fidelity $F_I$ of the identity operation, this fidelity
is also minimal for the error distribution shown in table \ref{table3}.
Specifically,
\begin{equation}
F_C \geq
 1-(\eta_z(\mbox{C})+\eta_z(\mbox{T})+\eta_x(\mbox{C})+\eta_x(\mbox{T}))
= 0.792.
\end{equation}
We therefore obtain an improved estimate of the entanglement capability
of our gate,
\begin{equation}
C \geq 2 F_C - 1 \geq 0.584.
\end{equation}
The more detailed analysis of the error distribution has thus provided
us with additional information on the entanglement capability of
our experimental device.

Finally, we can also improve our estimate of the entanglement
discrimination $D$ by considering the fidelity $F_D$ of the operation
that converts Bell state inputs into local $XZ$ eigenstates.
In this case, the errors that preserve the correct output states
are $\hat{F}_i=\{II,XI,IZ,XZ\}$,
corresponding to a Bell analyzer fidelity of
\begin{equation}
F_D=\chi_{\mbox{00,00}} +\chi_{\mbox{C0,C0}}
   +\chi_{\mbox{0T,0T}} +\chi_{\mbox{CT,CT}}.
\end{equation}
Again, the minimal fidelity can be obtained using the diagonal
matrix elements shown in table \ref{table3}, and the corresponding
minimal fidelity estimate is given by
\begin{equation}
F_D \geq
 1-(\eta_z(\mbox{T})+\eta_z(\mbox{B})+\eta_x(\mbox{C})+\eta_x(\mbox{B}))
= 0.806.
\end{equation}
Interestingly, this fidelity is a little bit higher than the fidelity
$F_C$ for entanglement generation. We therefore obtain a minimal
entanglement discrimination $D$ that exceeds the minimal entanglement
capability $C$ obtained from the same data,
\begin{equation}
D \geq 2 F_D - 1 \geq 0.612.
\end{equation}
The error analysis of the local $Z$ and $X$ operations thus
shows that our gate can successfully generate and distinguish
entangled states, with somewhat stronger evidence for the
reliability of Bell state discrimination.

\section{Conclusions}

As the analysis of the errors in our experimental quantum
controlled-NOT gate has shown, the classical logic operations
observed in a pair of complementary basis sets can provide
surprisingly detailed information about the performance of
a quantum device. In particular, it is possible to obtain
good estimates of the process fidelity $F_{qp}$, the
entanglement capability $C$, and the entanglement
discrimination $D$ from only a small fraction of the data
needed for a complete reconstruction of the process matrix
by quantum process tomography.
Interestingly, the complementary processes of the quantum
controlled-NOT are completely local. It is therefore possible
to estimate the non-local properties of the gate described
by the entanglement capability $C$ and the entanglement
discrimination $D$ without ever generating entangled states.

Besides the obvious advantages of gaining quick and efficient
access to the most important measures characterizing a
quantum process, the analysis of the process matrix in terms
of its observable effects also allows us to take a peek
inside the "black box" that is postulated in so many approaches
to quantum computation. In particular, it is possible to
identify the features of quantum coherence and entanglement
more directly with the experimentally accessible data
by identifying mathematical expansions that fit the specific
features of the quantum process under investigation.
Hopefully, this is only a first step towards a better understanding
of the still somewhat mysterious nature of quantum information
processes.

\section*{Acknowledgments}

This work was supported in part by the CREST program of the
Japanese Science and Technology Agency, JST, and the Grant-in-Aid
program of the Japanese Society for the Promotion of Science, JSPS.


\end{document}